\documentclass[fleqn,twoside]{article}
\usepackage{espcrc2}

\usepackage{graphicx}
\usepackage[figuresright]{rotating}

\newcommand{\AmS}{{\protect\the\textfont2
  A\kern-.1667em\lower.5ex\hbox{M}\kern-.125emS}}

\title{Ultrahigh Energy Cosmic Rays and Neutrinos}

\author{Todor Stanev\address{Bartol Research Foundation, 
        Department of Physics and Astronomy,
        University of Delaware, Newark, DE 19716, U.S.A.}}
\begin{document}

\begin{abstract}
 We discuss the relation between the highest energy cosmic rays
 (UHECR) and UHE neutrinos. The neutrinos produced in the sources
 of optically thin astrophysical sources have been linked to
 the UHECR emissivity of the Universe. The fluxes of cosmogenic
 neutrinos, generated in propagation by UHECR, also reflect the
 acceleration of these particles, the maximum acceleration energy,
 and the cosmological evolution of their sources. 
\vspace{1pc}
\end{abstract}

\maketitle

\section{Introduction}

  It is now a time when both subjects in the title are 
 of high scientific interest in the field of particle
 astrophysics. The interest is related to the emergence
 of new set of experiments that are obviously of much 
 better quality than what we know from the past.

  The Southern Auger Observatory~\cite{Auger} in Argentina is
 almost fully completed and is operating during construction.
 The Telescope Array~\cite{TA} (TA) is under construction in
 Utah. Both these giant air shower arrays are designed and 
 constructed to operate in hybrid mode, i.e. they employ both
 methods of air shower detection: surface array plus 
 optical detectors that follow the fluorescent light generated
 by the giant air shower in the atmosphere. The optical 
 detectors operate about 10\% of the time but the energy 
 assignment by integration over the shower longitudinal
 profile seems to be less model dependent than the classical method 
 of relating the primary energy to the shower particle density 
 on the ground. The latter depends much more on the hadronic 
 interaction model used in Monte Carlo studies of the shower
 development. The fraction of showers detected in hybrid mode
 by both detectors can be used for a better calibration of
 the surface array. The sizes of the surface arrays, 3,000 km$^2$
 for Auger and 1,000 km$^2$ for TA are very impressive, more
 than one order of magnitude higher than the previous largest
 array - Agasa.

  There are also big news on the neutrino front. The IceCube
 neutrino observatory~\cite{IceCube} on the South Pole is in
 the middle of  successful construction. Twenty two of the
 eighty underice strings are deployed and the deployment is
 now faster than initially expected. The Antares~\cite{Antares}
 detector in the Mediterranean is also successfully close to
 completion and the km3net~\cite{km3net} is working on the
 design of a cubic kilometer under water neutrino telescope 
 in the Northern hemisphere. With these cubic kilometer 
 neutrino detectors we approach for first time the dimensions
 needed for detection of astrophysical neutrinos - a gigaton 
 of target matter or 6$\times$10$^{38}$ target nucleons.

  Experiments of this size and quality can certainly make 
 the objects of their studies fashionable. The first 
 high statistic results of the Auger observatory were
 posted on ArXive.org and reported at the 20th International
 Cosmic Ray Conference in Merida, Mexico in front of a huge
 audience. Auger has set~\cite{Auger_gam} a strict limit 
 of 2\% on the gamma-ray contribution to the UHECR flux
 above 10$^{19}$ eV and in the following we shall assume 
 that the highest energy cosmic rays originate in 
 acceleration in powerful astrophysical objects.

 IceCube has not yet detected any astrophysical
 neutrino signals but the limits already obtained have let 
 to revisions of many models.
     
\section{Relation and differences between UHE cosmic rays and neutrinos}

 The connection between ultrahigh energy cosmic rays (UHECR)
 and neutrinos was first emphasized by Waxman\&Bahcall~\cite{WB}.
 This paper used an estimate of the cosmic ray emissivity of
 the Universe~\cite{W95} above 10$^{19}$ eV to calculate the
 maximum neutrino fluxes generated at the acceleration sites 
 of UHECR. The calculation has to assume an energy spectrum
 for these particles. It was taken to be a flat $E^{-2}$ 
 acceleration spectrum. Accounting for the cosmological evolution
 of the UHECR sources (and assuming that the same acceleration
 spectrum is followed to much lower energy) the maximum isotropic
 neutrino flux was calculated to be $E^2 dF/dE$ = 5$\times$10$^{-8}$
 per cm$^2$.ster.s and was called the {\em upper bound} of the
 isotropic neutrino fluxes. This bound applies to optically thin
 sources where the accelerated nuclei can leave the astrophysical
 source and does not limit the neutrino production in optically
 thick ones. 

\begin{figure}[htb]
\vspace*{-8pt}
\includegraphics[width=0.99\linewidth]{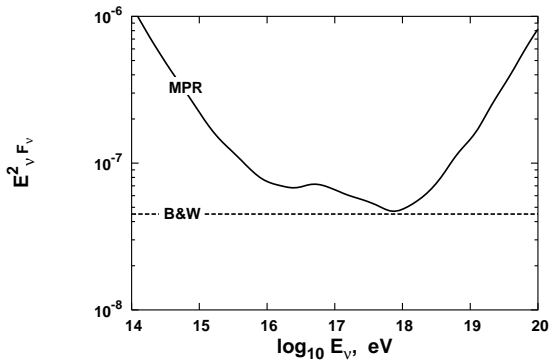}
\vspace*{-20pt}
\caption{The upper bounds of astrophysical 
 neutrinos calculated by Waxman\&Bahcall (straight line) 
 and Mannheim, Protheroe \& Rachen.}
 \vspace*{-10pt}
\label{WBMPR}
\end{figure}

 A better calculation emphasizing the simplifications in this
 approach was published by Mannheim, Protheroe \& Rachen~\cite{MPR}.
 Their {\em upper bound} touches the W\&B one at E$_\nu$ = 10$^{18}$ eV
 and is higher than it at all other neutrino energies. A comparison 
 between the two calculations is shown in Fig.~\ref{WBMPR}.
 Neither of these calculation include the neutrinos that are generated 
 by the UHECR on propagation to us - the {\em cosmogenic} neutrinos.
 
 Ultrahigh energy cosmic rays are not well understood for two reasons:
 it is indeed very difficult to appreciate how charged particles
 can be accelerated to energies as high as 10$^{20}$ eV in any
 astrophysical object, and because of their high energy loss
 in interactions on the microwave background (MBR) and other 
 universal photon fields, such as the infrared/optical background
 (IRB). Although we are discussing interactions of particles 
 exceeding 10$^{19}$ eV in the Lab the physics involved is very
 well known. Because of the low energy of the photon background 
 fields the center of mass energy is in the 1-10 GeV range and
 the cross sections are well known from accelerator experiments.
 There are two important processes: photoproduction interactions
 and $e^+ e^-$ pair production. The photoproduction process that
 generates the GZK~\cite{GZK} effect has in the MBR a threshold
 of about 3$\times$10$^{19}$ eV for protons. There are lower
 energy proton interactions in the IRB but they do not seem to be 
 important for the proton spectrum evolution in propagation.
 The minimum interaction length is at proton energy of
 5$\times$10$^{20}$ eV and is below 4 Mpc. At higher energy the
 energy loss length is about 14 Mpc, which for Hubble constant
 $h_0$ = 0.75 corresponds to redshift $z$=0.0035. A large fraction 
 of the proton energy loss in photoproduction interactions
 goes into neutrinos.
 Every time a charged pion is produced three neutrinos 
 (muon neutrino and antineutrino and an electron neutrino)
 are produced.

 At lower energy (smaller CM energy needed) the main energy loss
 process is the electron-positron pair production. This process
 has a threshold of 2$\times$10$^{18}$ eV. The cross section
 increases with proton energy, but the energy loss per interaction
 decreases. The combination leads to a minimum energy loss distance
 of 1.2 Gpc ($z$ = 0.3) at about 2$\times$10$^{19}$ eV. 
 Berezinsky\&Grigorieva~\cite{BerGri} first discussed the importance of
 this process for the UHECR spectrum after propagation.

 At energies lower than 2$\times$10$^{18}$ eV the main energy
 loss process is the adiabatic energy loss from the expansion 
 of the Universe. The energy loss length for $h_0$ = 0.75 is
 4 Gpc.   

\begin{figure}[htb]
\includegraphics[width=0.99\linewidth]{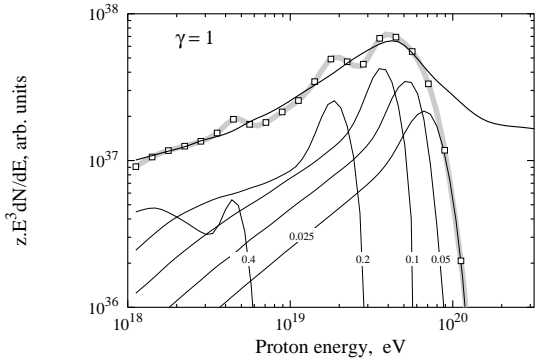}
\vspace*{-20pt}
\caption{Evolution of the cosmic ray spectrum in
 propagation on different distances.}
\vspace*{-10pt}
\label{pprop1}
\end{figure}

 Both the UHECR spectra from isotropically distributed
 cosmic ray sources and the neutrinos generated by them 
 depend on the cosmic ray acceleration spectrum and on the 
 maximum acceleration energy. There is, however, a big 
 difference related to the very different energy loss of
 cosmic rays and neutrinos. Neutrinos only suffer 
 adiabatic energy loss and easily propagate to us 
 from all redshifts while protons of arbitrary energy can 
 propagate from large redshifts. Figure~\ref{pprop1}
 shows the proton spectrum after propagation at 
 different redshifts, from 0.025 to 0.4
 The maximum energy in this graph is 10$^{22}$ eV with 
 exponential cutoff at 10$^{21.5}$ eV. Still, after 
 propagation on $z$ = 0.4, no protons of energy above
 10$^{19}$ remain in the cosmic ray flux. This fact 
 shows that the influence of the cosmological 
 evolution of the UHECR sources is not very significant
 for the highest energy cosmic rays.

\begin{figure}[htb]
\includegraphics[width=0.99\linewidth]{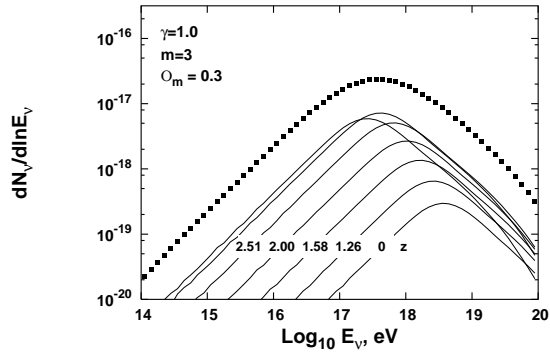}
\vspace*{-20pt}
\caption{Contribution of different redshifts to the 
 flux of cosmogenic neutrinos. One can see the adiabatic
 energy loss of neutrinos as the peak of the distributions
 change with redshift.}
\label{cosm1}
\vspace*{-10pt}
\end{figure}

 The cosmogenic neutrino fluxes, on the other hand, are
 very sensitive to the cosmological evolution of the sources.
 Figure~\ref{cosm1} shows a similar graph for cosmogenic 
 neutrinos. The contribution of different redshifts is 
 shown on a logarithmic scale of $z$ for the same 
 cosmological evolution model as in Fig.~\ref{pprop1} - 
 $(1+z)^3$ to $z$ = 1.9 and then constant to 2.8 with 
 exponential decline at higher redshift. Because of the small
 energy loss of neutrinos the contribution continues growing 
 after redshift of 2.5. Without cosmological evolution of
 the sources the highest contribution would have come from
 the contemporary Universe. 
 If we succeed in the detection of cosmogenic neutrinos we
 can understand the cosmological evolution of the extragalactic
 cosmic ray sources and create a better model of the highest
 energy cosmic rays~\cite{SS05}.
  
\section{Ultrahigh Energy Cosmic Rays}

\begin{figure}[htb]
\includegraphics[width=0.99\linewidth]{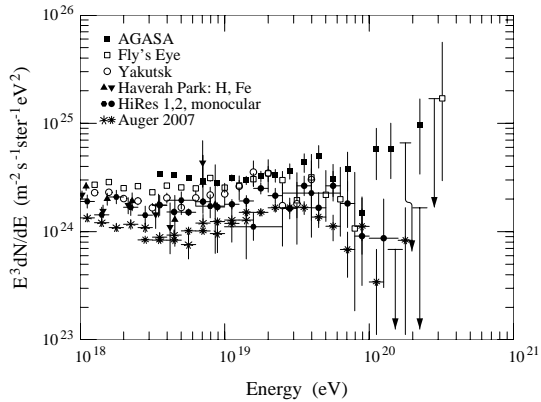}
\vspace*{-20pt}
\caption{Spectrum of the highest energy cosmic rays detected
 by different experiments.}
\label{fig807}
\vspace*{-10pt}
\end{figure}

 In this section we will present the newest set of data,
 compare it to older ones and to some of the available
 models. Figure~\ref{fig807} shows the data from all 
 experiments. The new Auger data come from the
 surface array normalized to the fluorescent detector
 energy assignment in hybrid events~\cite{mroth}. At lower
 energy only showers detected in hybrid mode are included~\cite{lperrone}. 
 Another energy spectrum, consistent with the shown ones,
 was derived from inclined showers.
 At 10$^{19}$ eV the difference in the energy 
 assignment between the highest and lowest flux
 (Agasa~\cite{AGASA} and Auger) is about 40\%. Auger supports the 
 measurement of HiRes~\cite{HiRes1,HiRes2} that shows
 a strong decrease of the cosmic ray flux above 5-6$\times$10$^{19}$ eV.
 The total number of events higher than 10$^{20}$ eV is two.
 These two data sets change our expectations for such events -
 it appears that expect to see 0.5 event in 1,000 km$^2$.ster.yr
 of exposure, about 10 times less than the Agasa estimate.
 The Auger energy assignment seems to be somewhat lower than that 
 of HiRes, maybe by about 20\% averaged over the whole energy range 
 shown. There are also some minor differences in the exact shape
 of the spectrum: the dip at 10$^{18.5}$ is better pronounced and
 the recovery at higher energy is faster. The GZK effect seems to 
 take place at slightly lower energy and the flux decrease is not
 as steep as in HiRes data set. As far as the highest energies 
 are concerned the shower statistics is too small to be analyzed.

 The question now is: which of the available models fit the experimental 
 spectrum the best. There are in principle three available models.
 The one of Berezinsky {\em et al}.~\cite{Ber1,Ber2} uses steep 
 acceleration spectrum (E$^{-2.7}$) down to about 10$^{18}$ eV.
 This model emphasizes the dip in the spectrum that is due to the
 $e^+e^-$ pair production loss in propagation and its conversion
 to purely adiabatic energy loss. This model is unique because it
 fits the cosmic ray flux above 10$^{18}$ eV with extragalactic
 cosmic ray protons  without any galactic component. The model fits
 the HiRes spectrum better than it fits the Auger one. 

 The second, also proton model, was suggested by Waxman\&Bahcall~\cite{WB02}
 and supported also by other authors.
 It uses a flat E$^{-2}$ acceleration spectrum, strong cosmological
 evolution of the cosmic ray sources and predicts a dip where the 
 flux of extragalactic cosmic rays intersects the galactic 
 cosmic ray component. This model requires that our Galaxy accelerates
 some cosmic rays to energies above 10$^{19}$ eV. 

 A third model uses cosmic rays at their sources with a composition 
 similar to the GeV galactic cosmic rays~\cite{Allard1,Allard2}.
 Since there is a significant fraction of heavy cosmic ray nuclei
 the energy loss in propagation is different in this model.
 Heavy nuclei lose energy in photodisintegration in the photon 
 fields. The energy threshold coincides with the giant dipole 
 resonance, and is thus much lower than photoproduction. The 
 nuclei lose nucleons in the process - the energy per nucleon is
 stable but the total energy per nucleus decreases. The total 
 energy loss length is not dissimilar to that for proton 
 photoproduction. As well as the other two models it fits the
 measured spectra quite well with an intermediate acceleration 
 spectrum of E$^{-2.2-2.3}$.

 The three models predict quite different nuclear compositions 
 for the UHECR. In the Berezinsky {\em et al} model the transition from
 galactic to extragalactic cosmic rays, and from very heavy to
 very light composition, is at the approach of 10$^{18}$ eV.
 In the flat spectrum model this transition is at significantly
 higher energy and UHECR should contain some iron nuclei even 
 above 10$^{19}$ eV. The mixed composition model is somewhat
 intermediate. In the whole energy range there is a significant
 fraction of protons released in the photodisintegration process
 complementing the primary and secondary nuclei. Only in the
 highest energy range the composition becomes light.

\begin{figure}[htb]
\vspace*{-10pt}
\includegraphics[width=0.99\linewidth]{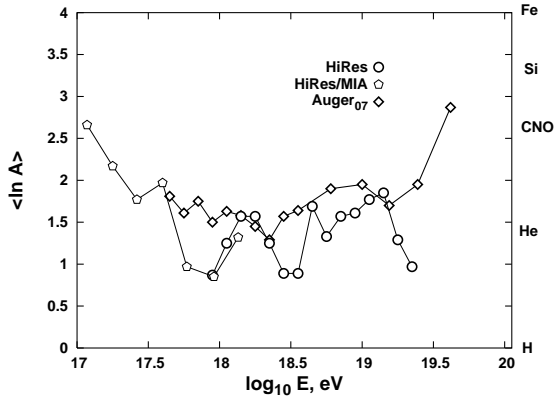}
\vspace*{-20pt}
\caption{Cosmic ray nuclear composition at the highest energies.
 The measured quantity is the depth of maximum which is converted
 by me to \protect$\langle lnA \rangle$.}
\label{comp}
\vspace*{-10pt}
\end{figure}

 The experimental data shown in Fig.~\ref{comp} do not seem
 to support any of the models.
 The circles and the pentagons show the HiRes and HiRes/MIA 
 data~\cite{HiRes_comp} on the shower depth of maximum converted
 to $\langle ln A \rangle$ using the Sibyll2.1 hadronic interaction
 model. The diamonds show the the Auger data~\cite{munger} treated
 the same way. 
 The composition measurement of Auger shows generally a heavier 
 composition, consistent with Helium. Excluding the highest
 energy points the differences between Auger and HiRes are not
 huge. The beautiful picture presented in Ref.~\cite{HiRes_comp}
 does not exist any more, and the composition picture seems to
 be much more complicated.

 The composition studies done by Auger (and probably HiRes)
 are hurt by the lack of understanding of the air shower 
 development revealed by the hybrid detection method. 
 If Auger did not have a fluorescent detector and used the
 surface shower densities as Agasa did, they would be two 
 changes in the current results: the energy estimate would 
 be higher by about 25\% and the cosmic ray composition
 would appear much heavier~\cite{rengel}. The currently used 
 hadronic interaction models are not able to explain this 
 effect. 

 One important result from the studies of the composition of
 UHECR is the strict limit of the fraction of $\gamma$-rays
 which is set to 2\% for all particles above 10$^{19}$ eV~\cite{Aug_gam}.
 This limit shows that at least 98\% of the UHECR particles are
 nuclei accelerated in astrophysical objects. 

\section{Cosmogenic neutrinos}

 Cosmogenic neutrinos are the neutrinos generated in photoproduction
 interactions of the propagating cosmic rays in the photon fields
 of the Universe. In the MBR the current threshold energy for proton
 photoproduction is about 3$\times$10$^{19}$ eV. Protons lose
 less than 20\% of their energy in the threshold energy range.
 We can roughly estimate the average cosmogenic neutrino energy
 as $\langle E_\nu \rangle \; =
 \; E_p \times K_{inel} / 4$ where $K_{inel}$ is
 the average energy loss of the protons per interaction
 and each neutrino takes 1/4 of the pion energy,  i.e.
 $\langle E_p \rangle \; = \; {\rm 10}^{18}$ eV.
 Cosmogenic neutrinos were first  proposed by
 Berezinsky \& Zatsepin~\cite{BZ68} and have been since
 calculated many times, most recently in Ref.~\cite{ESS01}.
 Neutrinos generated at higher redshift have adiabatic losses.
 In addition the cosmological evolution of MBR makes
 possible the interactions of lower energy protons 
 so the average energy of the cosmogenic neutrinos after 
 integration in redshift is of the same order.
  
\begin{figure}[thb]
\vspace*{-8pt}
\includegraphics[width=0.48\textwidth]{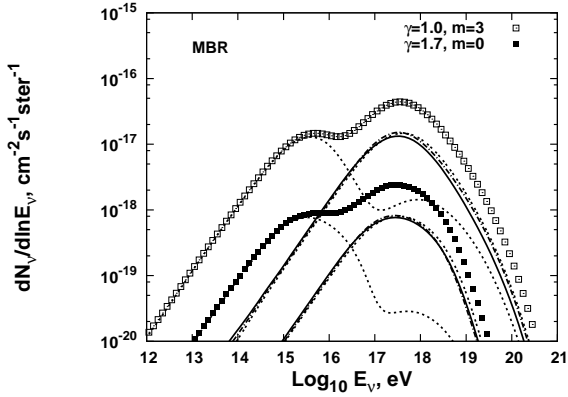}
\vspace*{-20pt}
\caption{Cosmogenic neutrinos generated by protons with a flat
 acceleration spectrum ($\gamma$=1.0) with 
 cosmological evolution and a steep one ($\gamma$=1.7) without
 evolution. Electron neutrinos are shown with a solid line,
 electron antineutrinos - with dots, muon neutrinos - with dashes
 and muon antineutrinos with dash-dotted line. The symbols 
 show the sum of all neutrino flavors. 
}
\label{allneu}
\vspace*{-10pt}
\end{figure}

 The spectrum of cosmogenic neutrinos depends on the UHECR 
 acceleration spectrum, the UHECR source distribution and very strongly
 on the cosmological evolution of the UHECR sources~\cite{SS05}.
 Flat acceleration ($\gamma$=1) models generate high flux
 because they contain higher number of interacting protons 
 of energy above 3$\times$10$^{19}$ eV and because they need
 strong cosmological evolution of the cosmic ray sources in order
 to fit the observed cosmic ray spectrum. The model of Berezinsky {\em et al},
 on the other hand, has much smaller number of  interacting
 protons for the same
 source emissivity and does not need cosmological evolution of the 
 sources. Figure~\ref{allneu} shows the fluxes of cosmogenic 
 neutrinos generated by these models.
 Note that the $\bar{\nu}_e$ spectrum peaks at about 10$^{15.3}$ eV
 while all other flavors peak at 10$^{18}$ eV. The reason is that
 $\bar{\nu}_e$ are generated in neutron decay rather than in 
 photoproduction interaction. They take a very small fraction of
 the neutron energy. Since the acceleration spectrum is protons 
 only, the neutrons are secondaries coming from charge exchange 
 interactions. Only at very high energy some secondary neutrons
 interact to produce high energy $\nu_e$ with the same distribution
 as the other flavors.

 It is of some importance to note that MBR is not the only 
 target for neutrino production. The second most important one 
 is the isotropic infrared and optical background (IRB). 
 Its number density is, or course, much lower, but lower 
 energy protons can interact in it and even in the case of
 flat acceleration spectra the number of interacting protons
 to a large extent compensates for the lower photon target 
 density.

\begin{figure}[thb]
\includegraphics[width=0.48\textwidth]{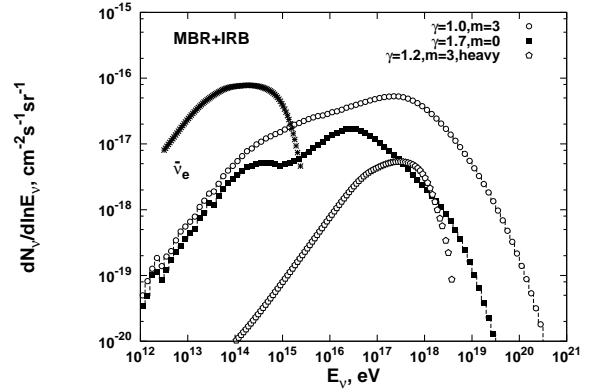}
\vspace*{-20pt}
\caption{Cosmogenic neutrinos generated by protons with a flat
 acceleration spectrum ($\gamma$=1.0) with 
 cosmological evolution and a steep one ($\gamma$=1.7) without
 evolution in MBR \& IRB. The pentagons show
 the spectrum of \protect$\nu_\mu + \bar{\nu}_\mu$ generated
 by the mixed composition model and the x's show the number
 of \protect$\bar{\nu}_e$ generated by the same model.
}
\label{all4+}
\vspace*{-10pt}
\end{figure}

 Fig.~\ref{all4+} shows the spectra of cosmogenic neutrinos 
 generated by a flat ($\gamma$=1.0) and steep ($\gamma$=1.7)
 UHECR acceleration spectra in the MBR and IRB. Since the
 steep injection spectrum has higher number of lower energy
 protons it provides more interactions in the IRB and decreases
 the difference between the two models. The flat acceleration 
 spectrum model contains more high energy cosmogenic neutrinos

 The mixed composition model generates mostly $\bar{\nu}_e$ 
 due to the decaying neutrons released by the photodisintegrating 
 neutrons~\cite{aveetal05}. Since there are many more such neutrons
 than  photoproduction interactions electron antineutrinos 
 dominate the cosmogenic neutrino flux. Electron neutrinos would 
 have the same spectrum as $\nu_\mu$ and $\bar{\nu}_\mu$, the 
 peak of which is similar to the neutrinos from the flat spectrum 
 model. Note that such much lower energy neutrinos have a
 significantly lower cross section and do not contribute much 
 to the total event rate. Their peak energy is also below the the
 Glashow resonance energy of 6$\times$10$^6$ GeV. Still, in case
 we are lucky enough  to detect cosmogenic neutrinos with the
 neutrino telescopes  under construction and design they could help
 a lot in  limiting the models for the origin of the ultrahigh
 energy cosmic rays. We have to remember, however, the expected
 event rate of cosmogenic neutrinos is small, less than 1 per
 km$^3$.yr and such detection requires new detector 
 technologies (radio detection?) that can cover hundred km$^2$.

\section{Summary}
  The new UHECR data sets agree that UHECR spectrum experiences 
 a steepening energy spectrum at about 5-6$\times$10$^{19}$ eV
 which looks consistent with the GZK effect which is due to 
 the energy loss of these particles in photoproduction or 
 photodisintegration interactions. It is not yet obvious 
 what are the parameters of the acceleration process and 
 what is the evolution of the cosmic ray sources. All these
 questions will probably have to wait until the total world
 statistics is increased by a large factor.

  The composition of all particles above 10$^{19}$ eV 
 is not yet established. Different experimental data 
 are not far away from each other, but do not currently
 contribute to the understanding of the UHECR origin.
 The advance in that respect is the strict limit of 2\%
 that the Auger collaboration has set on the fraction
 of $\gamma$-rays in the cosmic ray flux above 10$^{19}$ eV.
 
 Different UHECR models predict various fluxes of cosmogenic 
 neutrinos that are generated by the cosmic rays at 
 propagation from their sources to us. Detection of such 
 neutrinos and a comparison of their fluxes to the direct
 observations of UHECR would contribute significantly
 to understanding of the UHECR origin. For this to happen,
 however, we will have to rely on new, bigger neutrino
 observatories.
  
\section{Acknowledgments} 
 The author has benefited from discussions with
 P.~Blasi, D.~DeMarco, D.~Seckel, and A.A.~Watson.
 This work is supported in part by NASA APT grant NNG04GK86G.

\end{document}